\documentclass[17pt]{article}

\usepackage{amsmath,amssymb}
\usepackage{amsfonts}
\usepackage{eucal}
\usepackage{bigints}
\usepackage{todonotes}
\usepackage{graphics}
\usepackage[braket,qm]{qcircuit}
\usepackage{tikz}
\usepackage{caption}
\usetikzlibrary{arrows,tikzmark}
\usepackage[colorlinks=true]{hyperref}

\newcommand\blfootnote[1]{%
  \begingroup
  \renewcommand\thefootnote{}\footnote{#1}%
  \addtocounter{footnote}{-1}%
  \endgroup
}

\usepackage{authblk}
\title{New Directions in Quantum Music: concepts for a quantum keyboard and the sound of the Ising model}
\author[1]{Giuseppe Clemente}
\author[1,2]{Arianna Crippa}
\author[1]{Karl Jansen}
\author[1,2]{\newline Cenk T\"uys\"uz}
\affil[1]{\small Deutsches Elektronen-Synchrotron (DESY), Platanenallee 6, 15738 Zeuthen, Germany}
\affil[2]{\small Institut f\"ur Physik und IRIS Adlershof, Humboldt-Universit\"at zu Berlin, Zum Gro\ss en Windkanal 6, D-12489 Berlin, Germany}
\date{}

\begin{document}
\maketitle
\begin{abstract}
We explore ideas for generating sounds and eventually music by using quantum devices 
in the NISQ era using quantum circuits.
In particular, we first consider a concept for a ``qeyboard'', i.e.\ a quantum keyboard,
where the real-time behaviour of expectation values using a time evolving quantum circuit
can be associated to sound features like intensity, frequency and tone.

Then, we examine how these properties can be extracted from physical quantum systems,
taking the Ising model as an example. This can be realized by measuring physical quantities    
of the quantum states of the system, e.g.\ the energies and the magnetization obtained 
via variational quantum simulation techniques.
\blfootnote{Chapter submitted for publication in the book “Quantum Computer Music”, Edited by E. R. Miranda (Springer, 2022).}
    
\end{abstract}

\section{Introduction}
With the current acceleration in the development and improvement of quantum technologies,
it is conceivable that we shall witness an increasing influence of quantum ideas 
in everyday life, music included. Public availability and easy access to the Noisy Intermediate-Scale Quantum (NISQ)~\cite{Preskill2018} devices
allowed users of different backgrounds (e.g.\ composers, software developers, video game designers)
to experiment with them. Using principals of Quantum Mechanics to generate or manipulate
music is gaining popularity in the recent years~\cite{putz_quantum_2017,miranda_quantum_2020}.

In this chapter, we present two ideas. First, the Qeyboard, an attempt to turn 
a Quantum Computer into an instrument, allows the performer to exploit quantum effects such 
as superposition and entanglement. 
This opens up new avenues for live performances and musical compositions.
The second idea converts the algorithmic process of simulating a quantum system into music, 
making it possible to hear the properties of the quantum system as it evolves. 
Both of these ideas are simple enough to be practically realizable with the current technology
and little effort; they also allow for straightforward extensions and generalizations 
for later stages of this quantum revolution.

\section{Qeyboard: some concepts for a real-time quantum keyboard}
The idea of a \emph{quantum instrument}, i.e.\ 
a device capable of producing sounds as end-products of quantum mechanical processes 
instead of classical mechanical ones, 
is starting to take shape in the quantum music community~\cite{putz_quantum_2017}.
The ``true'' nature of the real world is quantum 
(at least, according to our most accurate description of nature), 
so any classical musical instrument is already quantum at the fundamental level, 
but the extent of quantum effects (such as superposition, interference, entanglement) 
is usually obscured by the macroscopic and incoherent nature of the phenomena involved
in the sound production.

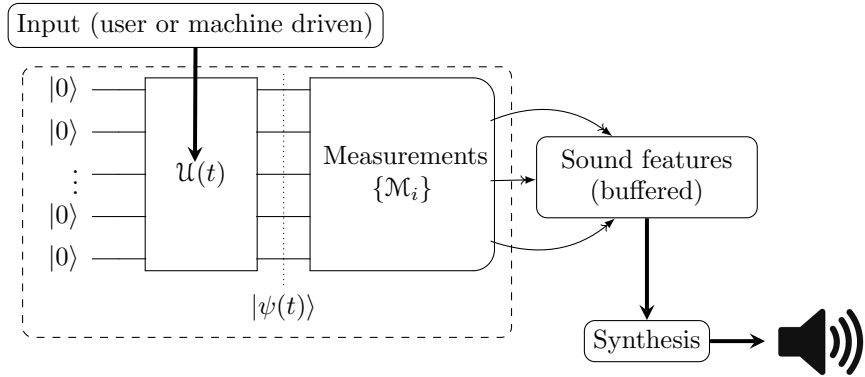
\begin{figure}[ht]
    \centering
    \tikzstyle{arrow} = [thick,->,>=stealth, line width=1.5pt]
    \begin{tikzpicture}[remember picture]
        \node (CIN) [draw, rounded corners] at (-1.30,0.2) {Input (user or machine driven)};
        \node (QCIRC) at (0,-2) {%
                $\begin{aligned}
            \Qcircuit @C=1em @R=.7em { 
                \lstick{\ket{0}} & \qw & \multigate{4}{\ \ {\mathcal{U}^{\subnode{Umarker}{}}(t)}\ \ }  & \qw \ar@{.}[]+<0.0em,0.6em>;[d]+<0.0em,-5.8em> &\multimeasureD{4}{\parbox{2.1cm}{\centering \text{Measurements}\\ $\{\mathcal{M}_i\}$}{\subnode{Mmarker}{}}}\\
              \lstick{\ket{0}} & \qw & \ghost{\ \ {\mathcal{U}(t)}\ \ } & \qw &\ghost{\parbox{2.1cm}{\centering \text{Measurements}\\ $\{\mathcal{M}_i\}$}} \\
              \lstick{ \vdots} & \qw & \ghost{\ \ {\mathcal{U}(t)}\ \ } & \qw &\ghost{\parbox{2.1cm}{\centering \text{Measurements}\\ $\{\mathcal{M}_i\}$}} \\
              \lstick{\ket{0}} & \qw & \ghost{\ \ {\mathcal{U}(t)}\ \ } & \qw &\ghost{\parbox{2.1cm}{\centering \text{Measurements}\\ $\{\mathcal{M}_i\}$}} \\
              \lstick{\ket{0}} & \qw & \ghost{\ \ {\mathcal{U}(t)}\ \ } & \qw &\ghost{\parbox{2.1cm}{\centering \text{Measurements}\\ $\{\mathcal{M}_i\}$}} \\ \\
                               & & & \ket{\psi{(t)}}}
\end{aligned}$};

        \node (SoundFeatures) [draw, rounded corners] at (4.7,-1.82) {\begin{tabular}{c} Sound features \\ (buffered)\end{tabular}};
        \node (Synthesis) [draw, rounded corners] at (4.7,-4) {Synthesis};
        \node (Speakers) at (7.0,-4) {\includegraphics[width=0.1\textwidth]{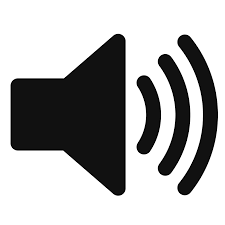}};

    \draw [arrow] (CIN) -- (Umarker);
    \draw [->latex] (Mmarker) -- (SoundFeatures);
    \draw [->latex, bend right] ([yshift=-0.8 cm]Mmarker.east) to (SoundFeatures);
    \draw [->latex, bend left] ([yshift=0.8 cm]Mmarker.east) to (SoundFeatures);
    \draw [arrow] (SoundFeatures) -- (Synthesis);
    \draw [arrow] (Synthesis) -- (Speakers);

    \draw [dashed, rounded corners] (-3.6,-0.35) rectangle (2.9,-3.95) ;
    \end{tikzpicture}
    \caption{Schematic pipeline diagram of a digital quantum instrument.
    The dashed box represents the part involving quantum processes: $\mathcal{U}(t)$ 
    is the circuit preparing the state $\ket{\psi(t)}=\mathcal{U}(t)\ket{0}$, 
    while $\{\mathcal{M}_i\}$ is a set of Hermitian operators representing measurements.}%
    \label{fig:digital_qinstrument_structure}
\end{figure}

In order to harness the full potentiality of quantum processes in the NISQ era, 
in the following discussion we will focus on the quantum circuit model as a 
convenient representation for abstract wavefunctions: 
a generic quantum state is prepared as a sequence of gates acting on qubits
starting from an initial fiducial state (usually the state with all qubits set to $0$).
Having prepared a specific wavefunction, one can then measure its properties
which can in turn be related to sound features to be classically synthetized,
according to the pipeline diagram depicted in Fig.~\ref{fig:digital_qinstrument_structure}.
 
In the following sections, we will describe some of the possible choices for each steps
of this pipeline, namely:
\begin{enumerate}
    \item real-time \textbf{interface} classical input and evolution of the circuit $\mathcal{U}(t)$ (and therefore $\ket{\psi(t)}$);
    \item set of \textbf{measurements} on the state $\ket{\psi(t)}$ at each frame;
    \item map between measurements (real or binary valued) 
        and sound \textbf{features};
    \item \textbf{synthesis} of the final sound from different sound sources. 
\end{enumerate}

\subsection{Real-time interface for evolving a dynamical parameterized quantum circuit}
The circuit structure, described abstractly in Fig.~\ref{fig:digital_qinstrument_structure} 
as a multi-qubit unitary operator $\mathcal{U}(t)$, 
is time dependent according to the time dependence of the input data.
In complete generality, the circuit represented by $\mathcal{U}(t)$ 
can include time-varying parameterized gates, as well as evolving topologies, where gates 
are added or removed at any time and at any point of the circuit.
Notice also that here we are not making any assumption on 
the continuity or differentiability of $\mathcal{U}(t)$ with respect to the time parameter $t$:
this makes possible abrupt changes in the properties of the wavefunction,
which in turns allows to model all possible shapes of 
ADSR envelopes (attach, decay, sustain and release). 

From the point of view of a human user, 
a good level of real-time manipulation of the circuit can be realized 
by pressing combinations of keys mapped to a finite set of single and double-qubit gates,
which are then added to the right end of the circuit.
These gates can be parameterless (like Pauli, Hadamard, CNOT or SWAP gates) or parameterized
like generic rotations; in turn, the parameters can be changed in real time 
using a slider (e.g.\ operated via mouse).
Another possibility, which requires some more work in terms of interface but 
enhances the level of manipulation expressibility, is to manage the whole circuit 
with a touch monitor where gates can be added or removed at specific points or
their parameters changed. This can be realized by simple gestures 
or even with multiple simultaneous action.

If one is interested instead in machine driven execution, 
the circuit dynamics can be represented by a predeterminated ``quantum music sheet'', 
or some other form of input, which drives the circuit changes in full generality
and without the limitation of the human user (limited pace and simultaneous actions).

\subsection{Measurements}
Here we discuss the second step in the pipeline of Fig.~\ref{fig:digital_qinstrument_structure},
namely the association between properties of the wavefunction produced by
measurements on the circuit $\mathcal{U}(t)$ and properties of the sound.
Notice that, in practice, it is not possible to run quantum circuits
and measurements continuously in $t$, so we would assume a reasonable sampling rate 
at discrete times $t_i$ 
which still allows us to capture the main features of the circuit dynamics without 
loss of expressibility.
The results of measurements can finally be interpolated during post-processing.

\subsubsection{Playing qubits as quantum strings}
In this paradigm, which we named ``quantum strings with counts to intensity'', 
we associate an oscillator with a specific frequency to any qubit register, 
while the corresponding sound intensity is determined by the average count of measurements 
with outcome $1$.
As a concrete example, we can consider a $q=8$ qubits system associated to 
the major scale in the octave $C4$ to $C5$, 
so that a circuit could be visually mapped to pentagram lines and spaces.
At every time $t$, one can make $n_{\text{shots}}$ measurements in all the qubit registers,
which produces in general different states in the computational basis in terms
of a dictionary containing the $0$-$1$ bitstring representation of the state and the
number of times that has been observed: 
$\{"00\dots 00":c_{00\dots 00}, "00\dots 01":c_{00\dots 01}, \dots \}$. 
The intensity associated to the $n$-th quantum string would then be determined 
by the ratio between the sum of counts of states with the $n$-th qubit register 
set to $1$ and the total number of shots:
\begin{equation}
    \mathcal{I}_n = \sum\limits_{\vec{s}\in \mathbb{Z}_2^q \vert s_n=1}\!\! \frac{c_{\vec{s}}}{n_{\text{shots}}} \; \; \in [0,1].
\end{equation}
In this way, in absence of noise and with a trivial circuit $\mathcal{U}(t_i)=I$, 
only the state with all qubits set to zero would be observed for every shot 
($c_{\vec{0}}=n_{\text{shots}}$), so the sound would be silence. 
A generic (non-diagonal) circuit $\mathcal{U}(t_i)$ 
would instead be associated to a generic distribution of count ratios 
${c^i_{\vec{s}}}/{n_{\text{shots}}}$.
Notice that the addition of noise and a finite number of shots would always introduce
fluctuations in the intensity associated to each quantum string at neighbouring times 
$t_i$ and $t_{i+1}$, even if the circuit does not change.

\subsubsection{Expectation values to continuous sound properties}
In this paradigm, the whole wavefunction is associated to different properties of 
the final synthetized sound. Unlike the previous approach, we will not associate single qubit
registers to specific and predetermined frequencies, but instead we will characterize 
the properties of the sound at each time $t_i$ in terms of expectation values 
of Hermitian operators (observables).
This idea can be implemented in a complete general way, 
but we will give a simple concrete example.
Let us consider a two qubits system and two Hermitian operators 
\begin{equation}\label{eq:hermop_simplexample}
    H_{f} \equiv \frac{1}{2} (I-\sigma^X) \otimes I,\qquad  H_{i} \equiv \frac{1}{2} I \otimes (I-\sigma^Z).
\end{equation}
At any time $t$, after applying the circuit $\mathcal{U}(t)$, one would then
estimate the expectation value of the observables in Eq.~\eqref{eq:hermop_simplexample}:
$\expval{H_f}(t)\equiv \langle \psi(t) \lvert H_f \rvert \psi(t) \rangle$ 
and $\expval{H_i}(t)\equiv  \langle \psi(t) \lvert H_i \rvert \psi(t) \rangle$, 
which takes any value in the range $[0,1]$.
Fixing a conventional frequency range $[f_{0},f_{1}]$,
the value $\expval{H_f}(t)$ can be associated to continuous values of frequency 
in that range by the linear relation $f(t) = f_0 + (f_1-f_0) \expval{H_f}(t)$,
while intensity would simply be $i(t) = \expval{H_i}(t)$.
These time-dependent frequency and intensity properties of a single sound 
will be then synthetized, as discussed in the next section.
In the specific example of Eq.~\eqref{eq:hermop_simplexample}, 
the two properties are commuting and it is intuitive how the sound can be manipulated by 
an appropriate rotation in the first or second qubit register.
In general, one can associate pairs of Hermitian operators to 
the frequency and intensity properties for a certain number of sounds $N_S$;
for example, one can build a set of observables 
by considering all the possible combinations of tensor products of $I$, and 
\begin{equation}\label{eq:projop}
\Pi_j=\frac{1}{2}(I-\sigma^{j}), 
\end{equation}
where $\sigma^{j}$ are the Pauli matrices and 
$\Pi_j$ is the projector to the eigenstate of $\sigma^{j}$ with eigenvalue $1$.
This set of operators can in principle be used to make a tomography of the wavefunction $\psi(t)$
at any time $t$ in order to extract all possible information from it. 
This choice would give a wide range of expressivity, since the number of properties 
could scale as $4^{q}-1$, but we recommend the user to select just a few of them
or a meaningful combination.
We want to stress also that one could add more sophisticated sound properties, 
like tone, which would require a different preprocessing stage during synthesis.

\subsection{Synthesis}
In this section we will briefly describe how to process the collection of properties 
$\mathcal{P}(t_i) = {\{(f_s(t_i),i_s(t_i))\}}_{s=1}^{N_S}$ for each sound at any time $t_i$ 
collected during the measurement stage (previous section),
and synthetize them in order to obtain a single waveform.
This can be done using inharmonic additive synthesis~\cite{additivesynth},
since the properties of each sound source is generally time-dependent.
First of all, an interpolation step must be performed in order to make 
the sounds properties vary continuously with time $\mathcal{P}(t)$.
The interpolation can be linear or higher order with some smoothing factor.
The final waveform is then built as follows\footnote{The phase of an individual 
sound could be added as a further property, but this would not be perceptible to the
listener's ear.}
\begin{equation}\label{eq:synthesis}
    w(t) = \frac{1}{\mathcal{N}}\sum_{s=1}^{N_S} i_s(t) \sin\Big[2 \pi f_s(t) t \Big],
\end{equation}
where a global normalization $\mathcal{N}$ has been added to make 
the waveform vary between $-1$ and $1$, so to avoid clipping effects.

\begin{figure}[h!]
    \centering
    \includegraphics[width=1.0\linewidth]{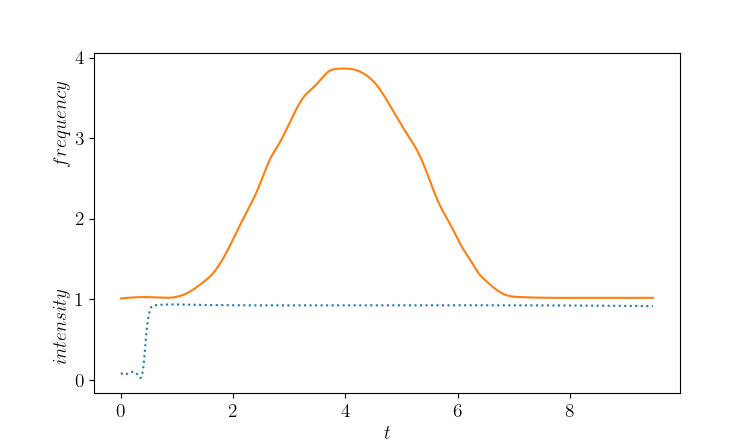}
    \caption{Simple example of circuit synthesis of single sound evolution 
    as described in the text: measurements are taken at every $0.1$ seconds, 
    quantum noise is present and a smoothed interpolation is applied at the post-processing stage,
with sampling rate $44100$ Hz.}%
    \label{fig:sheet_class2}
\end{figure}
\begin{figure}[h!]
    \centering
    \includegraphics[width=0.94\linewidth]{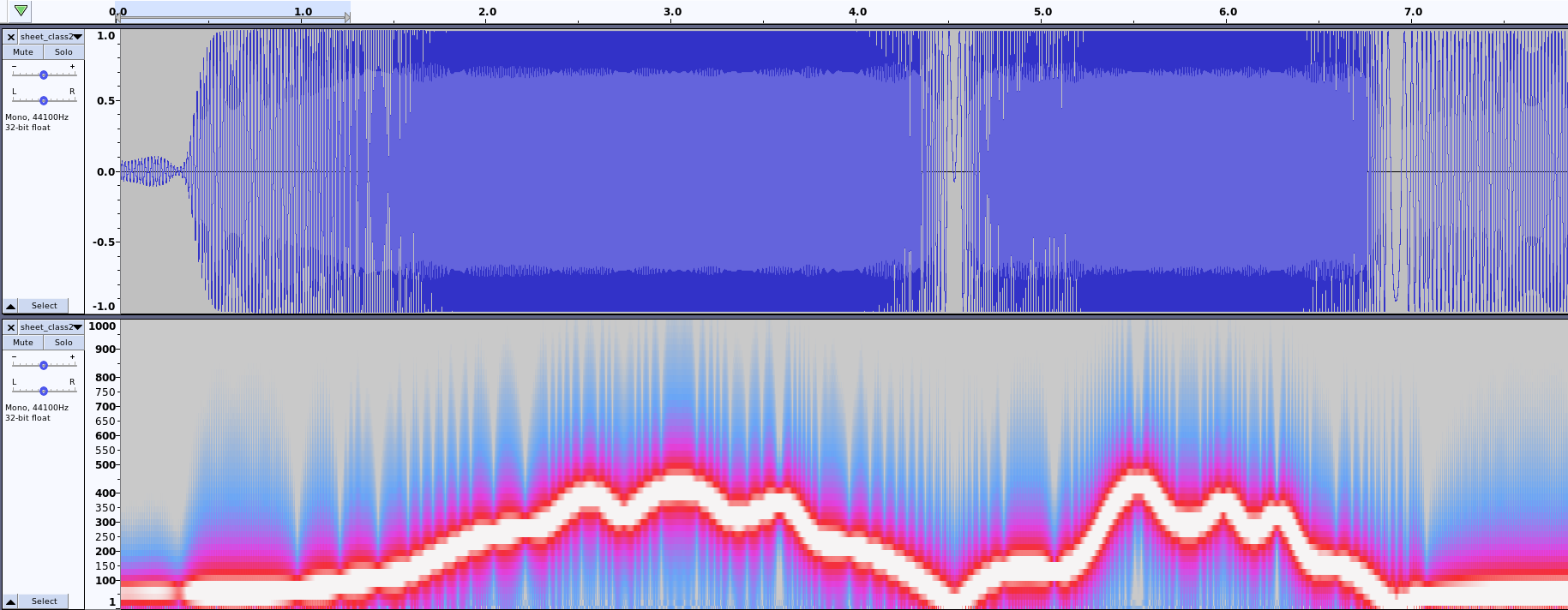}
    \caption{Inharmonic synthesis of the waveform obtained by using Eq.~\eqref{eq:synthesis} 
    with the frequency and intensity functions $f(t)$ and $i(t)$ shown in 
    Fig.~\ref{fig:sheet_class2},  with a base frequency $f_0=523.26$ and $f(t) \equiv f_0 \cdot \expval{H_f}(t)$. 
    Due to the presence of sampling noise, and since the synthesis is inharmonic, 
the spectrogram does not reflect straightforwardly the behaviour in Fig.~\ref{fig:sheet_class2}.}
    \label{fig:sheet_class2_sgram}
\end{figure}
As a first example, we consider a three qubits system,
with the following observables associated to intensity and frequency (fixed for the whole run):
\begin{align}
    H_i &= I \otimes I \otimes \Pi_Z \\
    H_f &= \big( 2 \Pi_Z \otimes I + I \otimes \Pi_Z + I\otimes I \big) \otimes I,
\end{align}
where $\Pi_i$ are the projection operators, defined in Eq.~\eqref{eq:projop}.
In this case, the intensity can be easily controlled by acting on the rightmost qubit register,
for example by applying $\sigma^Y$ or $\sigma^X$, 
while the first two registers from the left are associated
to a frequency, which can take values from $1$ to $4$.
\begin{figure}[ht!]
    \centering
    \includegraphics[width=1.0\linewidth]{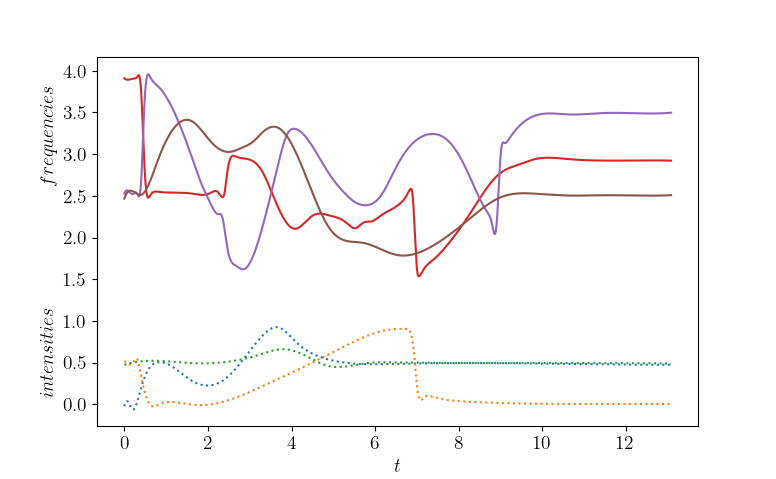}
    \caption{More complex synthesis for three sounds played simultaneously 
    with both evolving frequencies and intensities with some smoothing applied and synthetized 
    with sampling rate $44100$ Hz.}%
    \label{fig:sheet_test4}
\end{figure}
\begin{figure}[ht!]
    \centering
    \includegraphics[width=1.0\linewidth]{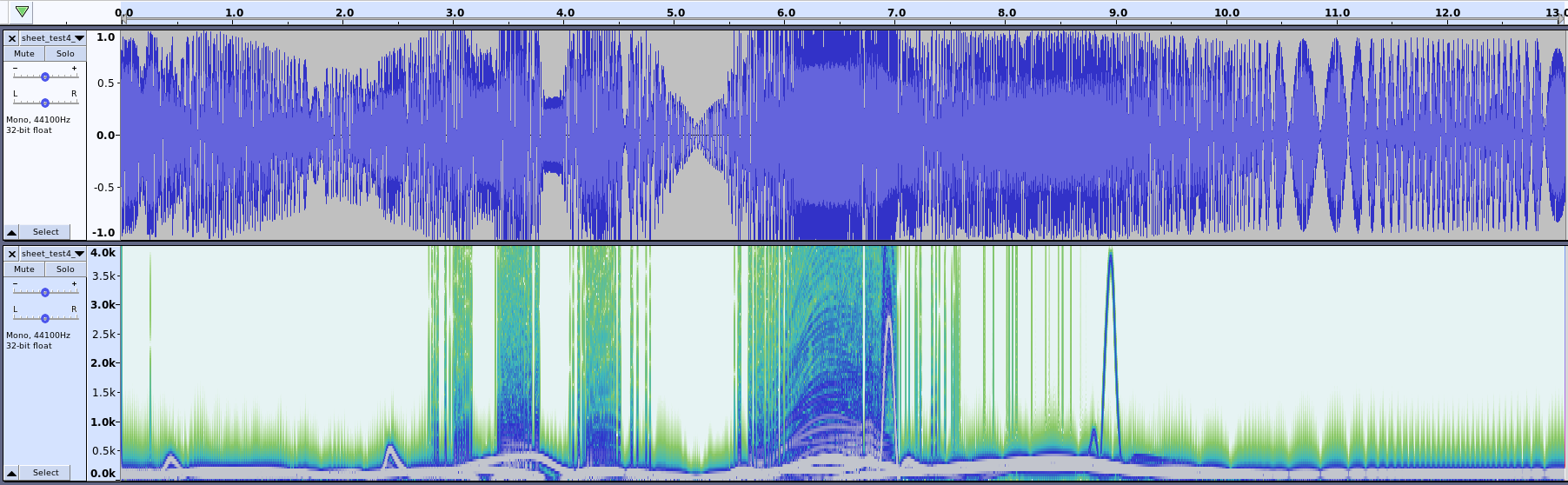}
    \caption{Inharmonic synthesis of the waveform obtained by using Eq.~\eqref{eq:synthesis} 
    with the frequency and intensity functions $f(t)$ and $i(t)$ shown in 
    Fig.~\ref{fig:sheet_test4},  with a base frequency $f_0=523.26$ and $f(t) \equiv f_0 \cdot \expval{H_f}(t)$. 
    Due to the presence of sampling noise, and since the synthesis is inharmonic, 
the spectrogram does not reflect straightforwardly the behaviour in Fig.~\ref{fig:sheet_test4}.}
    \label{fig:sheet_test4_sgram}
\end{figure}
For example, applying the following circuit evolution:
\begin{equation}
    \mathcal{U}(t) = \begin{cases}
                    I\otimes I \otimes I, & \text{ if } t < 0.5 s \\
                    I \otimes I \otimes \sigma^X, & \text{ if } 0.5 s < t < 1.0 s \\
                    e^{-i\frac{\pi (t-1 s)}{6 s} \sigma^X \otimes \sigma^X } \otimes \sigma^X, & \text{ if } 1.0 s < t < 4.0 s \\
                    e^{-i\frac{\pi (7 s-t)}{6 s} \sigma^Y \otimes \sigma^Y } \otimes \sigma^X, & \text{ if } 4.0 s < t < 7.0 s \\
                    I \otimes I \otimes \sigma^X, & \text{ if } t > 7.0 s,
    \end{cases}
\end{equation}
the behaviour in Fig.~\ref{fig:sheet_class2} is produced, where measurements are interpolated 
and then synthetized as in Fig.~\ref{fig:sheet_class2_sgram}.

Fig.~\ref{fig:sheet_test4} and~\ref{fig:sheet_test4_sgram} 
show another example of qeyboard dynamics, again for a
three qubits system, but with $6$ (non mutually commuting) Hermitian operators associated 
to intensities and frequencies for three sounds.
The qeyboard-driven circuit evolution and the set of observables 
used in this case are more complicated, so we will not put the specific details 
of its generation here, but it should be nevertheless possible to appreciate the degree of
customizability and expressivity which can be realized using this paradigm.

\section{The sound of the Ising model}
In this section we explore how to use physical systems to play quantum  music. 
To this end we will employ the spectrum 
and other properties of a quantum system; here we consider the \textit{Ising model}
as a convenient toy system.
The energies of the spectrum will be used as frequencies and other quantities 
such as the magnetization can be used for the intensites. 
These principles can be actually applied to many other physical systems, which 
would supply a very broad portfolio of sounds and eventually quantum music. 
The reason is that physical quantum systems can have very different properties 
showing a variety of phases and corresponding phase transitions. 

The Ising Model is a simple statistical mechanical system, demonstrating of it serving 
as a microscopic model for magnetism exhibiting a quantum phase transition 
from unmagnetized to a magnetized phase. 
It consists of discrete two-valued variables that represent the two possible states ($+1$ or $-1$) of 
magnetic dipole moments, or "spin". 
These spins are defined on a lattice and they interact with their nearest neighbours.
The Hamiltonian of the system has two terms
\begin{equation}
    H = - J\sum_{i}\sigma^Z_i\sigma^Z_j -h \sum_{i}\sigma^X_i,
\end{equation}
the first describes the interaction between neighbouring spins: 
if $J>0$, neighbouring spins prefer to be aligned 
($\uparrow\uparrow$ or $\downarrow\downarrow$), which denotes a \textit{ferromagnetic} behaviour. 
If $J<0$, the preferred combination is anti-aligned ($\uparrow\downarrow$), 
leading to an \textit{anti-ferromagnetic} behaviour.
The second term represents the action of an external magnetic field with amplitude $h$, 
which endows an energy advantage to the spins aligned to the magnetic field.
If the value of $h$ is sufficiently large, i.e. $h=O(1)$, the Ising model 
undergoes a phase transition where the fluctuations of the spins increase and interesting physics 
starts to happen. 
Here also the magnetization, defined in Sec.~\ref{magnetization}, decreases sharply as a 
function of the external magnetic field. 
It is the goal of this section to use the properties of the Ising system to 
generate also interesting quantum sounds and even quantum music.  

The idea is to use a variational approach, such as the Variational Quantum Deflation algorithm 
\cite{Higgott2019variationalquantum} (see Sec.~\ref{vqd}) to find 
pairs of eigenvalues and eigenvectors($\{(E_k,\ket{\psi_k}\}$, $k=0,1,\dots$) 
of the Ising Hamiltonian, and then convert the properties of the system into audible sounds.

\subsubsection{Variational Quantum Algorithms }\label{vqd}
\begin{figure}[h]
    \centerline{\includegraphics[width=5in, height=2.5in]{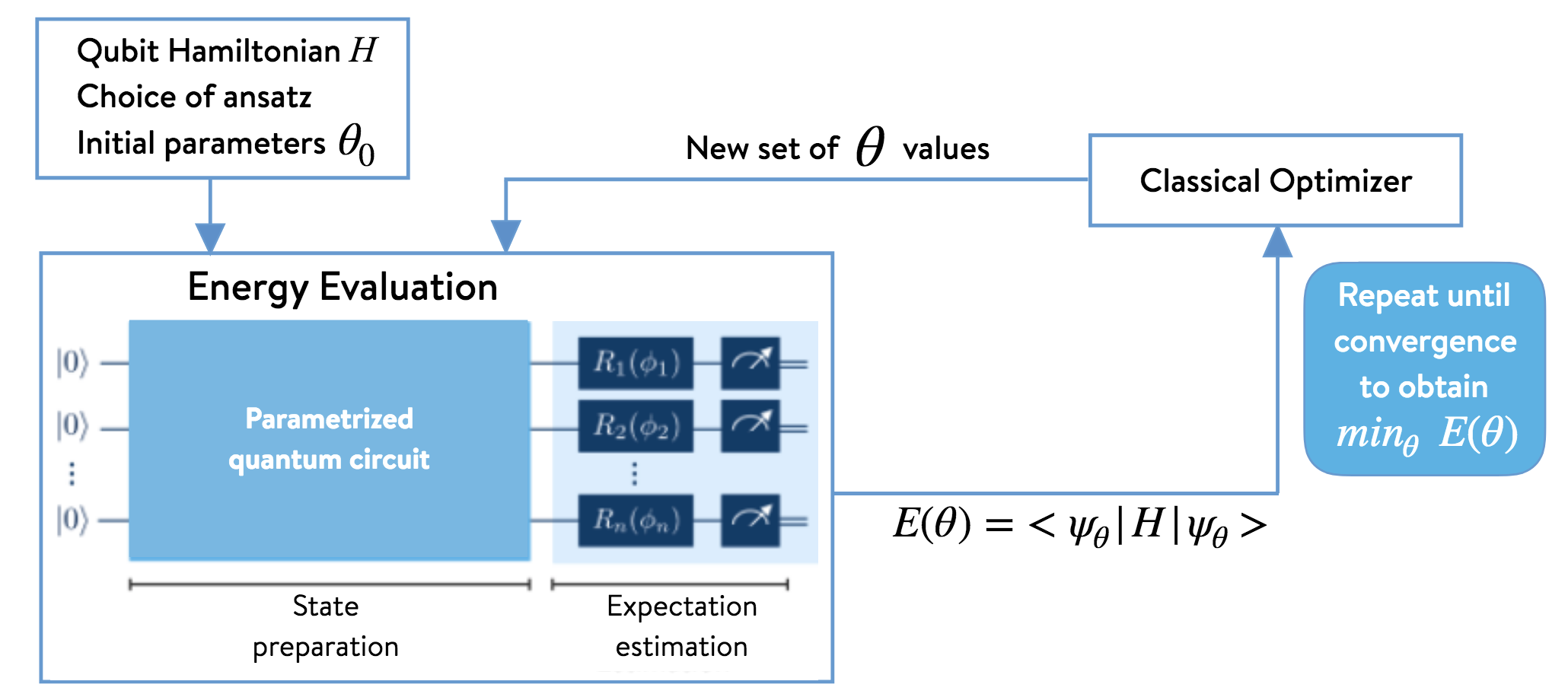}}
    \caption{Schematic procedure of the VQE algorithm. Source:~\cite{vqesource}.}
    \label{fig:vqe_tab}
\end{figure}

The Variational Quantum Eigensolver \cite{Peruzzo2014} 
uses a variational technique to find the minimum eigenvalue of the Hamiltonian of a given system.
An instance of VQE requires a trial state (ansatz), and one classical optimizer as
summarized in Fig.~\ref{fig:vqe_tab}. 

The ansatz is varied, via its set of parameters $\theta$, generating a state $\ket{\varPsi(\theta)}$ 
which allows to measure the energy as the expectation value of the Hamiltonian,
$\melem{\varPsi(\theta)}{H}{\varPsi(\theta)}$.
The classical optimizer then gives back a new set of parameters for the next 
computation of the energy. 
This procedure is repeated until convergence to the true minimum of the expection value 
is found. 

The VQE can be generalized also for computing excited states, 
for which the Variational Quantum Deflation (VQD) algorithm is used. 
The method has the following steps:
\begin{enumerate}
    \item Apply the VQE method and obtain optimal parameters $\theta_0^*$ and an approximate ground state 
        $\ket{\psi_0} \simeq \ket{\varPsi(\theta_0^*)}$.
    \item For the first excited state define a Hamiltonian:
        \begin{equation}
            H_1 = H + \beta \ket{\varPsi(\theta_0^*)} \bra{\varPsi(\theta_0^*)}
        \end{equation}
        where $\beta$ is a real-valued coefficient.
    \item Apply the VQE approach to find an approximate ground state of $H_1$.
    \item This procedure can be repeated for higher eigenstates. 
\end{enumerate}

\subsection{How to play a quantum system}
The aim of this section is to describe different possibilities for extracting sounds 
from a quantum physical system. By applying variational techniques we can get access 
to the observables of a quantum theory at the end of the minimization 
process and convert them to sounds. 
We can also measure the observables during the optimization itself and thus `play' 
quantum music running the VQE or the VQD algorithms.

Most of these techniques can be generalized to an arbitrary Hamiltonian, 
such as the one of Quantum Electrodynamics, or even more intricate systems from 
condensed matter or high energy physics. 

\subsubsection{Convert energy eigenvalues $E_k$ into frequencies}

\begin{figure}[h!]
    \centerline{\includegraphics[width=5in, height=3.5in]{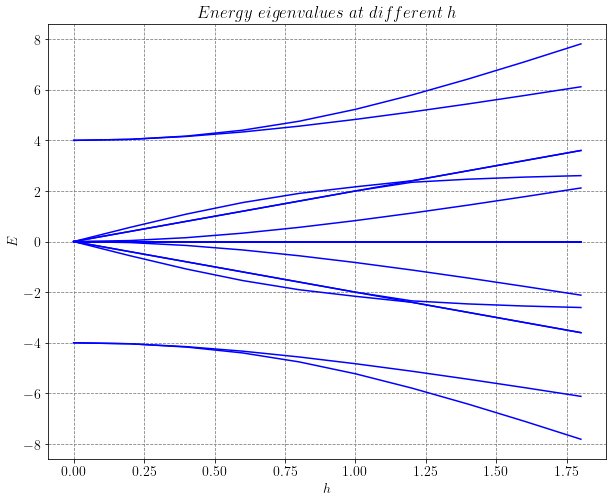}}
    \caption{The energy spectrum taken as frequencies and amplitudes $h$, as time, can be played as audible sounds.}
    \label{fig:eigsp}
\end{figure}

The first approach is to apply the VQD algorithm, compute the energy eigenvalues 
and convert them to audible frequencies.
To this end a suitable interval of the energies is chosen for a given 
value of the coupling $h$. Using $h$ as `time' variable, we can follow the behaviour 
of the corresponding frequencies and play them through an output device. 
Fig.\ref{fig:eigsp} shows the dynamics of the whole energy spectrum (16 eigenvalues) 
and can be naturally interpreted as a spectrogram.

\subsubsection{Use the callback results}

\begin{figure}[h]
    \centerline{\includegraphics[width=5in, height=3in]{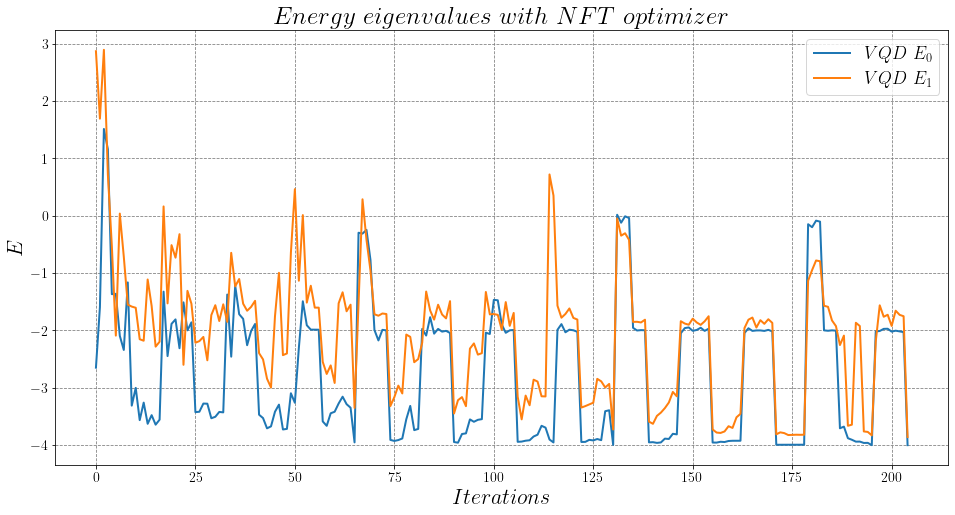}}
    \caption{Intermediate values during the optimization for ground state ($E_0$) 
        and first excited state ($E_1$) with NFT optimizer. }
    \label{fig:eigs}
\end{figure}

With this technique, the results for the ground state energy (or generic $E_k$) 
are collected during the VQD minimization with the NFT optimizer~\cite{PhysRevResearch.2.043158}.
The energies are measured now in each step of the optimization procedure and again 
converted into frequencies.
As can be seen in Fig.~\ref{fig:eigs}, the highly oscillatory behaviour 
of the energy values can be translated into frequencies. 
These oscillations are typical of the NFT algorithm but the detailed evolution depends
on the physical quantum system under consideration.  
Playing the frequencies of this hybrid quantum/classical approach 
can lead to very interesting sounds.

\subsubsection{Exploring how the sound changes across the phase diagram}
\label{magnetization}

With this method we include the \textit{magnetization} as an observable measured 
in the ground state $\psi_0$ and which is 
defined as 
\begin{equation}
    M=\frac{1}{N}\sum_i \melem{\psi_0}{\sigma^Z_i}{\psi_0}.
\label{eq:mag}
\end{equation}
 
\begin{figure}[h]
    \centerline{\includegraphics[width=5in, height=3in]{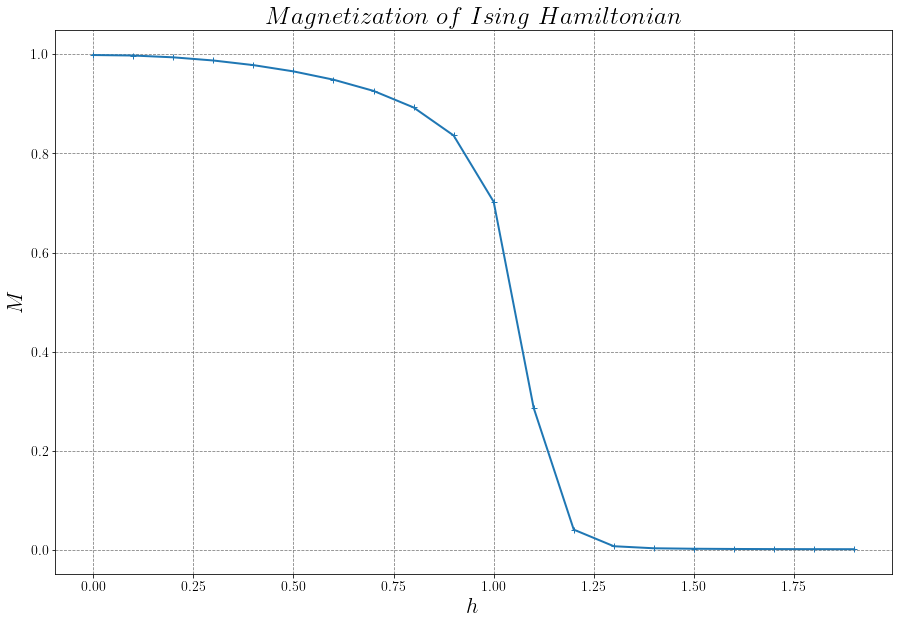}}
    \caption{Phase transition of Ising system. Varying the external magnetic field 
$h$, we go from a ferromagnetic phase to the paramagnetic phase by 
crossing a quantum phase transition. This can be clearly seen in the 
behaviour of the magnetization in Eq.~\protect{\eqref{eq:mag}} measured on the ground state.}
    \label{fig:mh}
\end{figure}

In Fig.~\ref{fig:mh} we can see that for small $h$ the magnetization is equal to one, 
this corresponds to a \textit{ferromagnetic}\footnote{Materials with a strong magnetization 
    in a magnetic field. They can turn into permanent magnets, 
i.e.\ have magnetization in the absence of an external field.} system. 
When $h$ increases, the magnetic term becomes more relevant and eventually the system reaches
a \textit{paramagnetic}\footnote{Materials with a weak induced magnetization in a magnetic field, 
which disappears when the external field is removed.} behaviour, with $M\sim 0$. 
In particular, we can observe a quantum phase transition when $h\sim1$.

The definition of the magnetization in Eq.~\eqref{eq:mag} can be generalized 
for higher excited states
\begin{equation}
    M_k=\frac{1}{N}\sum_i \melem{\psi_k}{\sigma^Z_i}{\psi_k} .
\label{eq:magk}
\end{equation}

These eigenstate-dependent magnetizations can be related to the intensity of the sound 
of the corresponding frequency, defined by the $k$-energy eigenvalue $E_k$. 

\section{Summary and outlook}
In this chapter we have discussed two approaches to generate sounds
through quantum devices based on the circuit model.
The first idea is to use the quantum computer as an instrument.
Here we use a real time evolution and manipulate the quantum circuit through a (quantum-)keyboard.
Measurements during the time evolution are performed to make them audible.
The ideas that we described for a quantum instrument are open for customization 
by the user at different stages of the pipeline and allow for a high degree of flexibility.

The second idea is to use quantum physical systems for generating sound and therefore 
to be able to actually listen to a true quantum system.
Here we followed the approach of assigning the role of `time variable' to some parameter 
of the model under consideration. In particular, in the case of the here discusses Ising model, the external 
magnetic field was chosen. Frequencies can then be computed from the energy eigenvalues and 
intensities from the magnetization measured in the corresponding eigenstates. 
In our opinion, 
using quantum systems to generate sound and eventually quantum music can lead to 
very interesting effects since such quantum models describe often  
very complex phenomena and phase diagrams with intricate physical properties. In addition, 
they exhibit phase transitions where large fluctuations can occur with strong 
correlations. We believe that these chracteristics of quantum systems can be harvest 
through the quantum mechanical principles of superposition and  
entanglement possibly leading even to new directions in music. 

Some resources (figures and sound files) for the examples that we discussed 
in the previous sections can be found in~\cite{qeybising}.

Our first step to generate quantum music presented in this chapter manipulates only
frequencies and intensities which can be obtained through measurements
of specific Hamiltonians or observables such as the magnetization.
Our approach can be generalized to also generate tones and even more complicated music properties.

\bibliography{bibl} 
\bibliographystyle{ieeetr}

\end{document}